\documentclass[12pt]{l4dc2020} 

\title[Teaching a robot to perceive the elapsed time - A reinforcement learning approach]{Teaching robots to perceive time - A reinforcement learning approach (Extended version)}
\usepackage{times}
\usepackage{multirow}
\usepackage{amsmath}
\usepackage{graphicx} % Enhanced LaTeX Graphics
\usepackage{algorithm}
\usepackage{algorithmic}
\usepackage{dcolumn}           % decimal-aligned tabular math columns
  
\newtheorem{problem}{Problem} 
\newtheorem{assumption}{Assumption}

\usepackage{enumitem}
\DeclareMathAlphabet\mathbfcal{OMS}{cmsy}{b}{n}
\usepackage{wrapfig}
\usepackage{caption}
\usepackage{subcaption}

\usepackage{tikz}
\usepackage{pgfplots}
\tikzset{every picture/.style={line width=0.75pt}} %set default line width to 0.75pt    
\pgfplotsset{compat=newest}
\usepgfplotslibrary{patchplots}
\usepackage{pgfplotstable}

\pgfplotsset{width=7cm,compat=1.13}
\usepgfplotslibrary{statistics}
  
\usepgfplotslibrary{fillbetween}
\usepackage{filecontents}

\pgfplotstableread[row sep=\\,col sep=&]{
    interval & carT \\

   0.3     & 1   \\
    0.6     & 1  \\    0.9   & 0 \\  
    0.3  & 3 \\
    0.6   & 7   \\    0.9  & 17 \\  1.2 & 33 \\ 1.5 & 84\\ 1.8 & 116 \\ 
    2.1 & 38 \\ 2.4 & 13 \\ 2.7 & 8 \\ 3 & 5 \\
    }\mydata
 
\def\FunctionF(#1){-0.18917*(#1)^3 + 0.84936*(#1)^2 - 0.54956*(#1) + 0.018702}%

\author{\Name{In\^es Louren\c{c}o} \Email{ineslo@kth.se} \\
 \addr Division of Decision and Control Systems, KTH Royal Institute of Technology, Stockholm, Sweden%
 \AND
 \Name{Bo Wahlberg} \Email{bo@kth.se}\\
  \addr Division of Decision and Control Systems, KTH Royal Institute of Technology, Stockholm, Sweden
  \AND
   \Name{Rodrigo Ventura} \Email{rodrigo.ventura@isr.tecnico.ulisboa.pt}\\
 \addr Institute for Systems and Robotics, Instituto Superior T\'{e}cnico, Lisboa, Portugal%
  }

\begin{document}
\maketitle

\addtolength{\abovedisplayskip}{-1.5mm}
\addtolength{\belowdisplayskip}{-1.5mm}
\addtolength{\textfloatsep}{-5mm}

%%%%%%%%%%%%%%%%%%%%%%%%%%%%%%%%%%%%%%%%%%%%%%%%%%%%%%%%%%%%%%%%%%%%%%
%     File: ExtendedAbstract_abstr.tex                               %
%     Tex Master: ExtendedAbstract.tex                               %
%                                                                    %
%     Author: Andre Calado Marta                                     %
%     Last modified : 2 Dez 2011                                     %
%%%%%%%%%%%%%%%%%%%%%%%%%%%%%%%%%%%%%%%%%%%%%%%%%%%%%%%%%%%%%%%%%%%%%%
% The abstract of should have less than 500 words.
% The keywords should be typed here (three to five keywords).
%%%%%%%%%%%%%%%%%%%%%%%%%%%%%%%%%%%%%%%%%%%%%%%%%%%%%%%%%%%%%%%%%%%%%%

%%
%% Abstract
%%
\begin{abstract}

Time perception is the phenomenological experience of time by an individual. %, present in every activity of our daily life.
%Sensory information has been proven to have an impact on the way we perceive the passage of time. 
In this paper, we study how to replicate neural mechanisms involved in time perception, allowing robots to take a step towards temporal cognition. 
Our framework follows a twofold biologically inspired approach.
The first step consists of estimating the passage of time from sensor measurements, since environmental stimuli influence the perception of time. Sensor data is modeled as Gaussian processes that represent the second-order statistics of the natural environment. The estimated elapsed time between two events is computed from the maximum likelihood estimate of the joint distribution of the data collected between them.
Moreover, exactly how time is encoded in the brain remains unknown, but there is strong evidence of the involvement of dopaminergic neurons in timing mechanisms. Since their phasic activity has a similar behavior to the reward prediction error of temporal-difference learning models, the latter are used to replicate this behavior. The second step of this approach consists therefore of applying the agent's estimate of the elapsed time in a reinforcement learning problem, where a feature representation called \textit{Microstimuli} is used.  
We validate our framework by applying it to an experiment that was originally conducted with mice, and conclude that a robot using this framework is able to reproduce the timing mechanisms of the animal's brain.  


%%
%% Keywords (max 5)
%%

%\begin{IEEEkeywords}
%temporal perception, reinforcement learning, gaussian processes, robotics, microstimuli.
%\end{IEEEkeywords}

%\noindent{{\bf Keywords:}} call centre, call scheduling, weapon and target assignment, dynamic programming, linear programming \\

\end{abstract}

%%%%%%%%%%%%%%%%%%%%%%%%%%%%%%%%%%%%%%%%%%%%%%%%%%%%%%%%%%%%%%%%%%%%%%
%     File: ExtendedAbstract_intro.tex                               %
%     Tex Master: ExtendedAbstract.tex                               %
%                                                                    %
%     Author: Andre Calado Marta                                     %
%     Last modified : 27 Dez 2011                                    %
%%%%%%%%%%%%%%%%%%%%%%%%%%%%%%%%%%%%%%%%%%%%%%%%%%%%%%%%%%%%%%%%%%%%%%
% State the objectives of the work and provide an adequate background,
% avoiding a detailed literature survey or a summary of the results.
%%%%%%%%%%%%%%%%%%%%%%%%%%%%%%%%%%%%%%%%%%%%%%%%%%%%%%%%%%%%%%%%%%%%%%

\section{Introduction}
\label{sec:L4DC_Introduction}
Our understanding of the human brain is still in its genesis. However, recent advances in robotics, machine learning and artificial inteligence research make it now possible to evaluate neuro-scientific theories $\emph{in silico}$. This consists of the reproduction of neural mechanisms by biologically inspired algorithms. In this work, we focus on the brain's timing mechanisms, which are responsible for $\emph{time perception}$ \citep{soares2016, Gouvea2015, Mauk2004}.
While the human and animal brain is known to have different areas responsible for temporal cognition, namely the basal ganglia, cyber-physical agents, such as robots, perform tasks based on algorithms that consider time as an external entity, such as a clock. 
The problem we thus study in this paper is:
%{\par\centering
\begin{center}
% Can a robot be able to perceive the passage of time like a human or animal?
\textit{How can biologically inspired mechanisms of time perception be reproduced in a robot?} %a non-biological system (e.g., a robot)? }
\end{center}
%\par}
This problem has been studied and motivated by works such as \citet{maniadakis2011}. In this paper, we take a step towards allowing the time representation of algorithms run by robots to be individually tailored and dynamically applied on its tasks. This might bring advantages not only at the level of validating neuro-scientific theories, but also of improving
the performance of robots on time-dependent tasks. We divide this goal in the two following steps.

Firstly, we emulate timing mechanisms algorithmically from environmental information. Successfully modelling this information can validate not only the idea that external stimuli influence the perception of time, but also that time can be estimated from raw information of the agent's sensors.

Secondly, an agent uses its estimate of the passage of time to perform tasks based on time perception mechanisms. Since the performance of the agent in these tasks will be different from that of an agent using the linear metric of time given by a clock, we can study whether this biologically inspired approach is advantageous for their performance on those tasks. For example, for the area of speech, having a perception of time would enable robots to learn to adapt their pauses in conversations to the situation and persons involved, as humans do.

The main contributions of this paper are the following:

\begin{itemize}[noitemsep, topsep=0pt]
	\item Presenting an approach to model environmental data collected from the sensors of a robot. 
	\item Under certain assumptions, showing that a correct time estimate can be obtained from data.
	\item Successfully applying mechanisms of temporal cognition in reinforcement learning problems.
	\item Empowering robots with the ability to replicate the actions of animals in time-dependent tasks. 
\end{itemize}

This paper is organized as follows. Section \ref{section:L4DC_Preliminaries} introduces preliminary background material and formulates the problem. Section \ref{section:Related work} presents related work that serves as a basis for this paper. Section \ref{section:our_framework} describes the framework used to solve the problem, with the main results being highlighted in Section \ref{section:L4DC_Numerical_results}. Finally, in Section \ref{section:L4DC_Conclusions}, conclusions are taken and future extensions are defined. 

%%%%%%%%%%%%%%%%%%%%%%%%%%%%%%%%%%%%%%%%%%%%%%%%%%%%%%%%%%%%%%%%%%%%%%
%     File: ExtendedAbstract_imple.tex                               %
%     Tex Master: ExtendedAbstract.tex                               %
%                                                                    %
%     Author: Andre Calado Marta                                     %
%     Last modified : 27 Dez 2011                                    %
%%%%%%%%%%%%%%%%%%%%%%%%%%%%%%%%%%%%%%%%%%%%%%%%%%%%%%%%%%%%%%%%%%%%%%
% A Calculation section represents a practical development
% from a theoretical basis.
%%%%%%%%%%%%%%%%%%%%%%%%%%%%%%%%%%%%%%%%%%%%%%%%%%%%%%%%%%%%%%%%%%%%%%

\section{Preliminaries}
\label{section:L4DC_Preliminaries}

In this section we formulate the setup of the problems we are considering in this paper.
We define the \emph{passage of time}, $\tau$, to be the time difference between two events. It is also denoted as \emph{elapsed time} or \emph{time interval}. Further, a \textit{time-dependent} task is one where time exerts an important role.
 
%\subsection{Notation}
%All vectors are column vectors, unless transposed. The \textit{i}th element of vector $v$ is $[v]_i$. Matrices are denoted by capital letters, and $p(\cdot)$ denotes the probability density.

\subsection{Problem formulation}
\label{ssection:problem_form}
% Explain why these two. why does answering these two questions gives us time perception?
To study the reproduction of biologically inspired time perception mechanisms in a robot, we create a twofold approach to solve the two main problems dealt with in this paper:

\begin{problem} A set of observations, $\mathcal{O}$, is collected during a certain interval of lenght $\tau$. Given the observations, estimate how long the interval was; that is, estimate $\tau$ given $\mathcal{O}$.
\label{problem1}
\end{problem}
An agent collects data $\mathcal{O} = \{y_t(i)\}_{i=1}^{M}$ from the environment, where $M$ is the number of sensors of the agent, and, from each, $N$ observations are uniformly taken throughout the interval $[0,\tau]$, at time instances $t_1, \dots, t_N$. These are represented as the $N$-dimensional vector $y_t(i) = [y_{t_1}(i), \dots,y_{t_N}(i)]^T$. 

How the estimate of the interval lenght can be used by the robot leads us to the second problem:

\begin{problem} Can a robot correctly perform a time-dependent task based on its estimate of $\tau$?
\label{problem2}
\end{problem}
To approach this problem, throughout this paper we consider the following discrete episodic reinforcement learning setup, modeled as a Markov decision process \citep{krishnamurthy2016}: at each timestep $t$ of each episode, the environment is in a certain state $s_t$, where $s \in \mathcal{S}$. The agent can perform an action, $a_t \in \mathcal{A}$, in the environment at each timestep, and, based on the \textit{value} of that action, it receives a reward $r_t$. Since we are interested in the \textit{value} of each action done by the agent, we consider a reinforcement learning setup with action selection using Q-values. 
The Q-value of a state-action pair, $Q(s,a)$, is defined as the \textit{expected return} from starting from state $s$ and performing action $a$. Hence, the actions are chosen with the goal of maximizing the expected return. 

\subsection{Experimental setup}
\label{ssection:experimental}
The time-dependent task we consider in this paper to validate our framework is the following:
using the framework described in Section \ref{ssection:problem_form}, a reinforcement learning agent is navigating around an environment and collecting data, $\mathcal{O}$, from its sensors. During each episode, the agent receives two stimuli, separated by a certain time interval, and estimates the duration of the interval (Problem \ref{problem1}). 

Since the state $s$ of the environment is influenced by the presence of stimuli, the interval duration estimated by the agent affects it. In other words, at the time that the agent estimates having seen the second stimulus, the state $s$ is changed to account for its presence. Since the state influences the agent's actions $a$, conclusions can be taken about the agent's time estimation mechanism (Problem \ref{problem2}) by analysing its sequence of actions.

\section{Related work}
\label{section:Related work}

This section presents the existing related biologically inspired frameworks that serve as a basis for approaching the two problems posed in Section \ref{ssection:problem_form}, one in each subsection.

\subsection{Time estimation from external stimuli (From Sensing to Time)}
\label{ssection:bayesian_framework}

Studies have shown that the way the passage of time is perceived in the human and animal brain is influenced by both neural internal mechanisms and innate estimation mechanisms from external stimuli. For example, \citet{BI} consider the Bayesian framework to estimate the elapsed time $\tau$ given the data $\mathcal{O}$:
\begin{equation}
p(\tau| \mathcal{O}) \propto  p(\mathcal{O}|\tau) \space p(\tau).
\label{eq:bayesian}
\end{equation}
The peak of the posterior distribution $p(\tau|\mathcal{O})$ is the Maximum \textit{a Posteriori} (MAP) and is considered to be the estimate of elapsed time, since it is the most likely time interval to have passed given the data.
The observations $\mathcal{O}$, collected from the environment, provide therefore an innate sensor-based estimate of the passage of time \citep{Eagleman2004, Brown1995}. They can be obtained from multiple sources of sensor data, such as images from a camera or time-series from a LIDAR. 

% 2. How can we model it?\\
The question is then \textit{how can this data be modelled?} 
One should take into account that environmental information does not change randomly, rather, it shows patterns of high correlation in both space and time \citep{dong1995}. Furthermore, to avoid handling excessive amounts of data, the observations are usually treated in a low-dimensional representation.
As an example, \citet{BI} studied how external stimuli introduce a bias on the perceived time, and considered the estimate of the elapsed time as a probabilistic expectation of stimulus change in the environment, that can be inferred from its second-order statistical properties. 
These are characterized by the mean $\mu$ and correlation between observations (such as points in a natural time-varying image). The latter is represented by the kernel $K$ and expresses how much the process changes from one timestep to the next, corresponding to the rhythm of change of the natural environment. They show that the power spectrum of the observations can be approximated by that of the Ornstein-Uhlenbeck (OU) function, which is a process of Brownian motion with friction \citep{OUcov1930}.
If the statistical properties remain constant in time, the process is stationary and thus the observations are modelled as stationary Gaussian processes with an OU kernel \citep{gaussian2007, Rasmussen2006}. 
%The MAP estimates are, therefore, equivalent to the ones given by an ideal observer of stationary Gaussian processes, whose equations are standard and found in .

\subsection{Temporal-difference learning and time perception (From Time to Action)}
\label{ssection:TDlearning}

Besides estimating time, it is important to replicate the way it is used by the brain in time-dependent tasks.
For long, dopamine neurons have been know to be involved in action selection and reward prediction mechanisms \citep{glimcher2011}. Besides these, they are also believed to be responsible for interval timing mechanisms, that is, to have the ability to encode the passage of time \citep{Gershman2014}. This is due to their phasic activity encoding a Reward-Prediction Error (RPE) signal, which is a biologically inspired signal that reflects the difference between the received and the expected reward on the basis of previous learning. This signal is present in Temporal-Difference (TD) learning models, which were introduced by \citet{sutton1990} and account for the need of a learning system that tries to predict the value of future events from the patterns of stimuli and rewards.
%\citet{Gouvea2015} confirmed that the speed with which the activity of certain neuronal populations change can predict the agent's judgement of interval duration. % In summary, the dopaminergic system can thus be seen as an internal clock of the brain and its activity can be reproduced using TD learning algorithms.
%
In TD learning algorithms with \textit{function approximation}, each state $s_t$ introduced in Section \ref{section:L4DC_Preliminaries} is described by a set of $D$ features, ${x_t(1),\dots, x_t(D)}$, that encode an animal's cognitive and sensory experiences (e.g., environmental stimuli) at each timestep $t$. 
The two most accepted theories nowadays to represent stimuli, given the need of consistency with what happens in the basal ganglia, are the \textit{Complete Serial Compound} \citep{montague1996, sutton1990} and the \textit{Microstimuli} \citep{Elliot2008}. Since the former presents inconsistencies in representing certain characteristics of the dopamine system \citep{daw2006}, the latter was proposed as an alternative representation. This feature representation is, to the best of our knowledge, the most consistent with basal ganglia mechanisms and is therefore used to solve Problem \ref{problem2} in the next section. %since neurons encode timestamps of different events but more recent time points are more precisely decoded than later ones.

\section{Replicating the perception of time using reinforcement learning}
\label{section:our_framework}

The following sections present the frameworks used to answer the problems posed in Section \ref{ssection:problem_form}, building upon the previous works mentioned in Section \ref{section:Related work}.
%\textit{Can a robot correctly use its estimate of the elapsed time to perform a time-dependent task?}

\subsection{From Sensing to Time}
\label{ssection:sensing_time}

To provide the sense of the passage of time to a cyber-physical agent, we adopt the approach of \citet{BI} to reformulate Problem \ref{problem1} as: \emph{given observations $\mathcal{O}$, compute the MAP estimate of \eqref{eq:bayesian} to find the $\tau$ that corresponds to the perceived elapsed time.}

\begin{assumption}[Uniform prior]
The prior distribution $p(\tau)$ is considered to be uniformly distributed so that the brain is assumed not to have \textit{a priori} information about the elapsed time. However, in the future, this prior can be integrated in the agent's architecture and adapted to the context of the task. 
\end{assumption}
Under uniform prior, the MAP in \eqref{eq:bayesian} coincides with the maximum likelihood estimate \citep{ljung1987}, which means that the estimate of the elapsed time is the maximum of $p(\mathcal{O}|\tau)$.
This corresponds to the probability of having observed $\mathcal{O} = \{y_i\}_{i=1}^{M}$ during the interval $\tau$, and is modelled as a zero-mean joint Gaussian distribution over the $N$ observations $y(0), \dots ,y(\tau)$ of all $M$ independent sensors: % 
\begin{equation}
p(y_{i}|\tau) = \mathcal{N}(y_i; 0, K_\theta) = \dfrac{1}{\sqrt{\det(2\pi K_\theta)}} e^{-\frac{1}{2}y_{i}K^{-1}_\theta y_{i}^T}.
\label{eq:equationBI} 
\end{equation}
This joint distribution has an unknown kernel function denoted by $K_\theta$, which is parametrized by $\theta$.

\noindent To solve Problem \ref{problem1} we first need to extend the work by \cite{BI} with a parameter estimation step to obtain the hyperparameters of the model from data: \emph{use Bayesian model selection to find the model's hyperparameters, $\theta$.}
For this, we maximize the logarithm of the likelihood in \eqref{eq:equationBI} with respect to $\theta$, which involves computing the respective derivatives:
\begin{equation}
\frac{\partial}{\partial\theta_j} \log p(y|\theta) = -\frac{1}{2} tr(\phi\phi^T - K^{-1}_\theta ) \frac{\partial K_\theta}{\partial\theta_j}, \text{ where } \phi = K^{-1} y
\label{eq:Gderivates}
\end{equation}

\begin{assumption}[Model selection]
Assume that, as justified in Section \ref{ssection:bayesian_framework}, the data comes from a Gaussian process with Ornstein-Uhlenbeck covariance.
\end{assumption}
In this case, the $N \times N$ OU kernel $K_\theta(\tau)$, where $\tau$ is the difference between time indexes, is: 
\begin{equation}
K_{\lambda,\sigma}(\tau) = e^{-\lambda |\tau|} + \sigma^2 \psi(\tau)
\label{eq:OU}
\end{equation}
% \text{, where } \delta(\tau) = \begin{cases}
%    1,& \text{if } \tau =0\\
%    0,              & \text{otherwise} \end{cases}.
Here, $\psi(0) =1$ and $\psi(\tau) =0$ for $\tau \neq 0$, and $\theta =[\lambda, \sigma]$ are the hyperparameters of the model.  Hence, Problem \ref{problem1} is solved by identifying the appropriate values for $\lambda$ and $\sigma$, in such a way that the properties of the Gaussian process approximate the ones of the data, and with these maximize \eqref{eq:equationBI}. The maximum $\tau$ is the estimate of the robot of the elapsed time between the two stimuli.

\subsection{From Time to Action}

The second part of our framework is called \textit{From Time to Action}, since we design a robot that, following the RL setup defined in Section  \ref{section:L4DC_Preliminaries}, can use its estimate of the elapsed time to correctly perform a sequence of actions \citep{Sutton2016}. 
Correctness entails similarly to the actions performed by an agent with temporal cognition on the same task.
The actions depend on the state sequence seen by the robot, which is influenced by the internal estimate of time between stimuli.
% 
%We denote the framework as ``\textit{Microstimuli}", and the specific elements of that framework as ``microstimuli".

In the simple case of linear \textit{function approximation} approaches, the Q-values referred to in Section \ref{ssection:problem_form} are defined as a weighted combination of (state and action dependent) features $x(s,a)$:
\begin{equation}
Q(s_t,a_t) = w^T x(s_t,a_t) = \sum_{j=1}^D w_t(j) x_t(j)  \text{, where, in this case, } x_{t}(j)= h_{t}  f\left(h_{t}, \frac{j}{m}, \beta\right).
\label{eq:Qfeatures}
\end{equation}
The weights ${w_t (1),..., w_t(D)}$ represent the strengths of the corticostriatal synapses and indicate how important each feature is for each state and action. Since we consider that the features are represented by the \textit{Microstimuli} framework, both cues and rewards, with a total of $\zeta$ per episode, deploy their own set of $m$ microstimuli. In total, there are $D =m \zeta$ microstimuli. Each one is given by the product of the trace height $h_{t}$, by basis functions $f$, according to \eqref{eq:Qfeatures}.
There, $x_{t}(j)$ is the level of each existing microstimulus at time $t$. Knowing how much a microstimulus has decayed due to its slowly decaying memory trace can be seen as a basis for the elapsed time. If $h_{t}$ decays exponentially and $f$ are Gaussian basis functions with center $\nu$ and width $\beta$, then $
h_{t} = \exp\{- (1-\xi)t\}$, and $f(h, \nu, \beta) = \frac{1}{\sqrt{2 \pi}} \exp\big\{ - \frac{(h - \nu)^2}{2 \beta^2}\big\}$, where $\xi$ is the decay parameter.
%This \textit{Microstimuli} feature representation therefore encodes a certain uncertainty in the temporal prediction, and, a
%As time goes by, different microstimuli become more or less active since later ones are wider, shorter and have a later peak.

The action selection mechanism used is the $\epsilon$-greedy, where $ a_{t}=\arg \underset{a}{\mathrm{\max } } Q(s_t,a)$ with probability $1-\epsilon_{t}$, and it is random otherwise. %The exploration parameter, $\epsilon_t$, decays according to $\Gamma$ as $\epsilon_{t} = \Gamma \epsilon_{t-1}$.
The TD learning algorithm introduced in Section \ref{ssection:TDlearning} is used to estimate the Q-values. The eligibility traces, $e_t$, are an essential attribute of reward learning that, when multiplied by the error, expand the influence of the presence of a state through time \citep{singh1996}. Furthermore, the error, $\delta_t$, is the difference between the expected and received reward $r_t$ at each timestep. Being $\gamma$ the discount rate and $\eta$ the decay parameter that determines the plasticity window of recent stimuli, the update equations of these parameters and the weights $w_t$ are:
\begin{equation}
\begin{aligned}
w_{t+1}(j) = w_t(j) + \alpha \delta_t e_t(j);
\delta_t = r_t + \gamma \max_{a} Q(s_{t+1},a)- Q(s_t,a_t);
e_{t+1}(j) = \gamma \eta e_t(j) + x_t(j).
\end{aligned}
\label{eq:Qweights}
\end{equation}

The estimate of the elapsed time, $\tau$, computed in Section \ref{ssection:sensing_time}, determines at which timestep a new set of microstimuli is deployed in the features $x(s,a)$, which influences the $Q$-value of the state-action pair and therefore the action $a$ chosen by the agent.
The framework is summarized in Algorithm \ref{alg:RL}.
Hence, Problem \ref{problem2} is solved by analysing the sequence of actions chosen by the agent in this TD learning framework. If it replicates the actions taken by a real animal in a time-dependent task, we claim that we were able to provide temporal cognition to the agent.

\begin{algorithm}[h]
	\caption{From sensing to action based on perceived time}
	\label{alg:RL}
	\begin{algorithmic}[1]
		%		\Require
		\STATE Initialize $Q(s,a) = 0$, for all $s \in \mathcal{S}$, $a \in \mathcal{A}$, and $w(1),\dots,w(D)$ randomly (e.g. $w(j) \in [0,1]$)
		\FOR{each episode}
			\STATE Initialize $s_0$
		\FOR{each timestep $t$}
			\IF{first stimulus == 1}
				\STATE Update $x_t(1), \dots ,x_t(D)$ according to \eqref{eq:Qfeatures}
				%\STATE (action should be random walk)
			\ELSIF{has received stimulus 1 but not 2}
				\STATE Collect data $y_t(1), \dots ,y_t(M)$
				%\STATE (action should be press button short/long)
			\ELSIF{second stimulus == 2}
				\STATE Compute the estimate of the elapsed time, $\tau$, by maximizing \eqref{eq:equationBI}
				\STATE Update $x_\tau(1), \dots ,x_\tau(D)$, according to \eqref{eq:Qfeatures}
				%\STATE (action should be press button short/long)		
			\ENDIF
			\STATE Choose $a_t$ from $s_t$ according to the Q-values. Take action $a_t$, observe $r_t$, $s_{t+1}$
			\STATE $\delta_t \gets r_t + \gamma \max_{a} Q(s_{t+1},a)- Q(s_t,a_t)$
			\STATE $e_{t+1}(j) \gets \gamma \eta e_{t}(j) + x_{t}(j)$
			\STATE $w_{t+1}(j) \gets w_{t}(j) + \alpha \delta_t e_{t}(j)$
			\STATE $s_t \gets s_{t+1}$
		\ENDFOR
		\STATE Until $s$ is terminal
		\ENDFOR
	\end{algorithmic}
\end{algorithm}

%Eligibility traces, denoted by $e$, are an essential feature of reward learning that, when multiplied by the error, expand the influence of the presence of a state through time \citep{BI}. We consider the update rule:
%\begin{equation}
%e_{t}(s,a)=\begin{cases}
%1+ \gamma \Lambda e_{t-1}(s,a) , & \text{if $s=s_{t}, a= a_{t}, Q_{t-1}(s_{t},a_{t}) = \underset{a}{\mathrm{\max }} Q_{t-1}(s_{t}, a)$ }\\
%0, & \text{if $Q_{t-1}(s_{t},a_{t}) \neq \underset{a}{\mathrm{\max }} Q_{t-1}(s_{t}, a)$} \\
%\gamma \Lambda e_{t-1}(s,a), & \text{otherwise}
%\end{cases},
%\label{eq:etraces}
%\end{equation}
%where $\gamma$ is the discount rate and $\Lambda$ the decay parameter that determines the plasticity window of recent stimuli.

%The action selection mechanism considered is the $\epsilon$-greedy, given by
%\begin{equation}
%a_{t}=\begin{cases}
%\arg \underset{a}{\mathrm{\max } } Q(s_t,a) , & \text{with probability $1-\epsilon_{t}$}\\
%\text{random action}, & \text{with probability $\epsilon_{t}$}
%\end{cases},
%\label{eq:action}
%\end{equation}
%where $\epsilon_t$ decays according to $\epsilon_{t} = \Gamma \epsilon_{t-1}$ and $\Gamma	$ is the decay parameter.
%
%

%\input{L4DC_RL}

\section{Case Study and Numerical Results}
\label{section:L4DC_Numerical_results}
In this section, we evaluate our complete framework described in Section \ref{section:our_framework} in a time-dependent task, with the goal of comparing the behavior of an agent using Algorithm \ref{alg:RL} with that of an animal. %, to conclude that the perception of time in the former resembles that of the latter.
%The simulations were done in a CPU ...

\subsection{Background}
\label{ssection:casestudy}
% Case study

\cite{soares2016} performed a time-dependent experiment with mice: a mouse can press three buttons: ``Start", ``Short" or ``Long". When it presses the former, two auditory cues, separated by a certain time interval that can be either short or long, are presented to the animal. Based on how much time it estimates to have passed between both cues, the animal then presses the button corresponding to its length (``Short" or ``Long"). 
If the action is correct, the animal is rewarded with water or food.

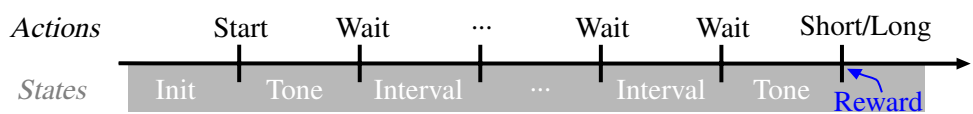
\begin{figure}[h]
\centering
   %\footnotesize
\begin{tikzpicture}[x=0.75pt,y=0.75pt,yscale=-0.7,xscale=0.8]
%uncomment if require: \path (0,83.22915649414062); %set diagram left start at 0, and has height of 83.22915649414062
%Shape: Rectangle [id:dp9003953698921725] 
\draw  [draw opacity=0][fill={rgb, 255:red, 0; green, 0; blue, 0 }  ,fill opacity=0.28 ] (102.17,40.96) -- (604.17,40.96) -- (604.17,75.96) -- (102.17,75.96) -- cycle ;
%Straight Lines [id:da014844376175189078] 
\draw [line width=1.5]    (96,40.96) -- (117.16,41.05) -- (625.17,41.23) (172,27.57) -- (171.99,54.57)(248,27.59) -- (247.99,54.59)(324,27.62) -- (323.99,54.62)(400,27.65) -- (399.99,54.65)(476,27.68) -- (475.99,54.68)(552,27.7) -- (551.99,54.7) ;
%Straight Lines [id:da6840012801056443] 
\draw    (96,40.96) -- (630.17,41.23) ;
\draw [shift={(633.17,41.23)}, rotate = 180.03] [fill={rgb, 255:red, 0; green, 0; blue, 0 }  ][line width=0.08]  [draw opacity=0] (8.93,-4.29) -- (0,0) -- (8.93,4.29) -- cycle    ;
%Curve Lines [id:da6950827697328705] 
\draw [color=blue  ,draw opacity=1 ]   (577.17,61.9) .. controls (573.33,48.46) and (575.94,54.98) .. (557.57,46.09) ;
\draw [shift={(555.17,44.9)}, rotate = 386.93] [fill=blue  ,fill opacity=1 ][line width=0.08]  [draw opacity=0] (8.93,-4.29) -- (0,0) -- (8.93,4.29) -- cycle    ;
% Text Node
\draw (57,15) node [xslant=0.25] [align=left] {Actions};
% Text Node
\draw (55,60) node [color={rgb, 255:red, 128; green, 128; blue, 128 }  ,opacity=1 ,xslant=0.25] [align=left] {States};
% Text Node
\draw (132,60) node [color={rgb, 255:red, 255; green, 255; blue, 255 }  ,opacity=1 ] [align=left] {Init};
% Text Node
\draw (173,15) node  [align=left] {Start};
% Text Node
\draw (207,60) node [color={rgb, 255:red, 255; green, 255; blue, 255 }  ,opacity=1 ] [align=left] {Tone};
% Text Node
\draw (285,60) node [color={rgb, 255:red, 255; green, 255; blue, 255 }  ,opacity=1 ] [align=left] {Interval};
% Text Node
\draw (438,60) node [color={rgb, 255:red, 255; green, 255; blue, 255 }  ,opacity=1 ] [align=left] {Interval};
% Text Node
\draw (250,15) node  [align=left] {Wait};
% Text Node
\draw (477,15) node  [align=left] {Wait};
% Text Node
\draw (514,60) node [color={rgb, 255:red, 255; green, 255; blue, 255 }  ,opacity=1 ] [align=left] {Tone};
% Text Node
\draw (401,15) node  [align=left] {Wait};
% Text Node
\draw (325,15) node  [align=left] {...};
% Text Node
\draw (362,60) node [color={rgb, 255:red, 255; green, 255; blue, 255 }  ,opacity=1 ] [align=left] {...};
% Text Node
\draw (567,15.5) node  [align=left] {Short/Long};
% Text Node
\draw (575,68.23) node [color= blue ,opacity=1 ] [align=left] {Reward};
\end{tikzpicture} 
\caption[RL task]{Desired state transition, obtained when the optimal action (on the top row) is chosen.}
\label{fig:RLtask}
\end{figure}

We replicate this experience in a simulated robot using the following environment: %Using a temporal-difference learning algorithm, the robot should use its estimate of the elapsed time to reach a certain goal. 
the state at which the environment is, at each timestep, is one of three: $\mathcal{S}=$\{Init, Tone, Interval\}, and the action the agent can perform is one of four: $\mathcal{A}=$\{Start, Wait, Short, Long\}. After pressing the \textit{Start} button, the state of the environment changes to \textit{Tone} and the number of \textit{Interval} states between the next \textit{Tone} state is randomly chosen. % and denoted as $I$, $I \in \mathbb{R}$. 
After the second \textit{Tone} state the agent should choose the action \textit{Short} or \textit{Long} that corresponds to that number. If the correct action is chosen, a positive reward is given to the agent. %The choice of the interval duration is done based on 50\% of probability of being short, and 50\% of being long. Inside the chosen interval, there is a uniform probability of choosing any value within that interval. 
The schematic representation of the optimal experiment is shown in Figure \ref{fig:RLtask}.%, where the black line represents the passage of time. 

%\begin{table}[]
%	\centering
%	\caption[Reinforcement learning time-dependent task]{Setup of the RL time-dependent task. The value of the intersection between a row, $s_{t}$, and a column, $a_{t}$, indicates the state $s_{t+1}$ for which the agent goes in the next timestep. N means that the agent gets a negative reward, and Y a positive, and in both cases the experiment end. In all the others the reward is neutral, $r_{t}=0$.}
%	\label{tab:RLtask}
%	\begin{tabular}{lllllll}
%		\cline{4-7}
%		\multicolumn{3}{l}{\multirow{2}{*}{}} & \multicolumn{4}{l}{Actions} \\
%		\multicolumn{3}{l}{} & Start & Wait & Short & Long \\ \cline{4-7} 
%		\multirow{3}{*}{\rotatebox[origin=c]{90}{States}} & 0) & \multicolumn{1}{l|}{Init} & 1 & 0 & N & N \\
%		& 1) & \multicolumn{1}{l|}{Sound} & N & 1 or 2 & N or Y & N or Y \\
%		& 2) & \multicolumn{1}{l|}{Interval} & N & 1 or 2 & N & N
%	\end{tabular}
%\end{table}

%{rgb, 255:red, 74; green, 131; blue, 226 }
%
\begin{wrapfigure}{r}{0.3\textwidth}
\centering
\begin{tikzpicture}
\begin{axis}[
	height=5cm,
	width=5cm,
	xmax = 25,
	ymax =25,
	xlabel={Simulated interval [s]},
	ylabel={Estimated interval [s]},
	grid style=dashed,
]
\addplot table[x=x,y=y] {data.dat};

\addplot [name path=upper,draw=none] table[x=x,y expr=\thisrow{y}+\thisrow{err}] {data.dat};
\addplot [name path=lower,draw=none] table[x=x,y expr=\thisrow{y}-\thisrow{err}] {data.dat};
\addplot [fill=blue!10] fill between[of=upper and lower];
\end{axis}
\end{tikzpicture}
\caption{\label{fig:time_estimate}Estimated $\tau$ for each interval duration.}
\end{wrapfigure}
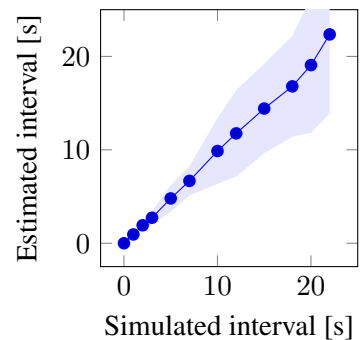
\subsection{Results and discussion}
\label{ssection:results}
In this section we present the main results obtained for the two problems discussed in this paper.

\vspace*{5px}
\textbf{Result 1}  \textit{ The elapsed time is successfully estimated.}
\vspace*{5px}

\noindent The first result obtained in this work concerns correctly computing the hyperparameters of the model. This means that following \eqref{eq:Gderivates} we can estimate the $\lambda$ and $\theta$ that maximize the likelihood function, being the estimated values similar to the original ones, as shown in Appendix \ref{appendix}. In this case, $y_i$ is considered to be the \textit{i}th angle of the simulated robot's LIDAR, from which data is collected while the robot does the \textit{Wait} action between tones.
After estimating the parameters of the collected data, we are able to estimate the elapsed time $\tau$. Figure \ref{fig:time_estimate} shows the estimated $\tau$ for different intervals, from which it can be concluded that our estimate is accurate for this range of intervals. This shows that Problem \ref{problem1} is correctly solved according to Section \ref{ssection:sensing_time} for this task.
It can be further concluded that the average estimated duration is not affected by the length of the interval to be estimated, but its standard deviation increases with the interval lenght. This is consistent with what happens with humans, since the longer the interval, the harder it is for us to estimate it \citep{sucala2011}.
%\begin{figure}[h]
%  \centering
%  \includegraphics[width=0.4\linewidth]{Figures/EstMean2.png}
%  \caption{Estimated $\tau$ for each real $\tau$.}
%  \label{fig:time_estimate}
%\end{figure}
 
\vspace*{5px}
\textbf{Result 2}  \textit{ The performance of the agent in the task is similar to that of the mice.}
\vspace*{5px}

\noindent A set of microstimuli $x$ are deployed in the state $s$ of the environment when the agent receives the first tone, and at the moment it estimates having heard the second. The agent performs the task in Section \ref{ssection:TDlearning} having these two sets of microstimuli affecting the states of the experiment in each episode. Its performance is evaluated according to how far it goes in the episode before making a wrong action, as is shown in Appendix \ref{appendix}. 
It was concluded that after the exploration phase is over, the agent has learned to perform the correct sequence of actions until receiving the second tone, and choosing the correct button if the interval was very short or very long. However, it can be seen in Figure \ref{fig:MS_8a} that intervals in the boundary are often misclassified, which is in accordance with the timing mechanisms of the brain, since for humans and animals these intervals are also harder to distinguish.
Figure \ref{fig:MS_8b} shows the corresponding psychometric curves, representing the probability of an interval being classified as long. The curve in red shows the original results of the paper \cite{soares2016} with the mice performing the experiment, which are similar to the ones of the agent, in blue. 
% xtick={0.6, 1.2, 1.8, 2.4, 3},
%0.3, 0.6, 0.9, 1.2, 1.5, 1.8, 2.1, 2.4, 2.7, 3
\begin{figure}[h]
\begin{minipage}[t]{0.45\textwidth}
\centering 
	\begin{tikzpicture}
    \begin{axis}[
            ybar,
            grid,
            height=4cm,
       		width=7cm,
            symbolic x coords={0.3, 0.6, 0.9, 1.2, 1.5, 1.8, 2.1, 2.4, 2.7, 3},
            xtick ={0.6, 1.2, 1.8, 2.4, 3},
            ylabel={Incorrect intervals},
            xlabel={Interval duration [s]},
            grid style=dashed,
        ]
        \addplot table[x=interval,y=carT]{\mydata};
    \end{axis}
	\end{tikzpicture} 
	\caption{Intervals incorrectly classified based on their corresponding duration. \\Total = 327, average = 1.65, median = 1.8.}
	\label{fig:MS_8a}
\end{minipage}
\hfill
\begin{minipage}[t]{0.45\textwidth}
\centering 
\begin{tikzpicture}
\definecolor{clr1}{RGB}{81,82,83}
\begin{axis}[
       % axis y line=center,
       % axis x line=middle, 
       % axis on top=true,
        xmin=0.5,
        xmax=2.5,
        ymin=0,
        ymax=1,
        height=4cm,
        width=6.5cm,
        grid,
        xtick={0.6,1,1.5,2, 2.4},
        ytick={0.1,0.5,1},
        ylabel={p(\textit{Long} choice) [\%]},
        xlabel={Interval duration [s]},
        legend pos=north west,
        grid style=dashed,
    ]
%\addplot [domain=-5:5, samples=50, mark=none, ultra thick, black]
 %{\FunctionF(x)};
% \node [left, blue] at (axis cs: 3.6,42) {$x^3-3x$};
 \addlegendentry{Mouse}
\addplot [
    color=red,
    mark=square,
    ultra thick,
    ]
    coordinates {
		(0.6,0.05)
		(1,0.1)
		(1.2,0.2)
		(1.35,0.35)
		(1.65,0.55)
		(1.8,0.72)
		(2,0.85)
		(2.4, 0.95) };
\addplot [
    color=blue,
    mark=triangle,
    ultra thick,
    ]
    coordinates {
		(0.6,0.01)
		(0.9,0.04)
		(1.2,0.12)
		(1.5,0.53)
		(1.8,0.82)
		(2.1,0.9)
		(2.4, 0.97) };
		\addlegendentry{Robot}
\end{axis}
\end{tikzpicture} 
\caption{Psycometric curves corresponding to the probability of an interval being classified as \textit{Long}.}
\label{fig:MS_8b}  
\end{minipage}  
\end{figure}

In summary, the main results that show the similarity in the behavior of mice and of the robot and that allow us to claim that the robot was able to obtain a time perception in this case similar to the humans and animals are the following:
\begin{itemize}[noitemsep,topsep=0pt]
    \item Environmental data effects the time estimate, and from the former, we can compute the latter;
    \item Uncertainty in the estimation of the interval duration increases with the length of the interval;
	\item The uncertainty in the classification of intervals is on the boundary between short and long;
	\item The error of the TD learning model behaves similarly to the firing rate of domapine neurons, decreasing with the reward expectancy. These results are shown in Appendix \ref{appendix}.
\end{itemize}

%%%%%%%%%%%%%%%%%%%%%%%%%%%%%%%%%%%%%%%%%%%%%%%%%%%%%%%%%%%%%%%%%%%%%%
%     File: ExtendedAbstract_concl.tex                               %
%     Tex Master: ExtendedAbstract.tex                               %
%                                                                    %
%     Author: Andre Calado Marta                                     %
%     Last modified : 27 Dez 2011                                    %
%%%%%%%%%%%%%%%%%%%%%%%%%%%%%%%%%%%%%%%%%%%%%%%%%%%%%%%%%%%%%%%%%%%%%%
% The main conclusions of the study presented in short form.
%%%%%%%%%%%%%%%%%%%%%%%%%%%%%%%%%%%%%%%%%%%%%%%%%%%%%%%%%%%%%%%%%%%%%%

\section{Conclusions}
\label{section:L4DC_Conclusions}

In this paper, we proposed an algorithm for providing temporal cognition to an agent. We exploited results from Bayesian inference to estimate the passage of time from data, and from TD learning feature representations to teach the agent to succeed in time-dependent tasks. Due to the choice of features representing time, we showed that in this context we are able to provide the agent a similar perception of the passage of time to the one that humans and animals have.
%Our results have implications in, \textbf{for example, ..., where ... }.
%We conclude that the robot learned to behave similarly to mice

In the future, this framework shall be implemented in a real robot. One interesting direction could be to map the movement of the robot to the statistical properties of the estimated processes. This would imply that time can be learnt disregarding whichever type of movement the robot performs.
%Furthermore, other methods than the maximization of the marginal likelihood could be used for the estimation of the hyperparameters, since this method does not guarantee global convergence, only local.
Furthermore, instead of assuming that the covariance function of the processes is Ornstein-Uhlenbeck, deep learning could be used to analyze the non-parametric distribution of the real data.
Finally, another future direction can be to include in the estimation of the subjective time the influence that affective states have, for variables such as attention or difficulty of the tasks performed.

% - A proper prior distribution of the internal estimate could be advantageous

%As for the use of the time estimate, However, a better could be to use for example a recurrent neural network instead of a Q-learning algorithm for learning, and compare this results as well.

%0.6	1	  1.2     1.35    1.5   1.65     1.8     2         2.4
%0.05	0.1    0.2    0.35     x     0.55    0.72   0.85       0.95

\appendix

\section{}
\label{appendix}
In this section we present supplementary information to the paper, including details of the implementation and additional results.

% Values given to the parameters
\subsection{ Simulated robot}
$M$ is the number of angles of the LIDAR and in our simulation environment $M=365$, meaning that we had environmental measurements from all directions around the robot.
However, in \eqref{eq:equationBI}, of those 365 usually around only 15 were needed to obtain a good time estimate.

\subsection{\textit{Microstimuli} framework}
In our environment described in Section \ref{ssection:experimental}, in each episode there are two cues (the two tones) and one reward, which means that $\zeta = 3$. We considered that each of these would deploy a set of $m=6$ microstimuli, which gives $D = 3 \cdot 6 = 18$.
The basis functions had width $\beta= 0.1$ and their decay parameter was $\xi= 0.9$.
% nPoints = 60; lenMax = 30, 

\subsection{Temporal-difference learning algorithm}
For the action selection mechanism with $\epsilon$-greedy, where
\begin{equation}
a_{t}=\begin{cases}
\arg \underset{a}{\mathrm{\max } } Q(s_t,a) , & \text{with probability $1-\epsilon_{t}$}\\
\text{random action}, & \text{with probability $\epsilon_{t}$}
\end{cases},
\label{eq:action}
\end{equation} 
the exploration parameter, $\epsilon_t$, had initial value $\epsilon_0 = 0.3$ and decayed according to $\epsilon_{t} = \Gamma \epsilon_{t-1}$, with decay parameter $\Gamma =	0.9995$.

In the algorithm update equation for the weights $w$, the error $\delta$ and the eligibility traces $e$, the discount rate was $\gamma= 0.1$, the decay was $\eta= 0.95$, and $\alpha = 0.2$.

\subsection{ Hyperparameters}
Table \ref{tab:hypEst} shows an example of the obtained parameters $\lambda$ and $\sigma$ of the kernel function $K_{\lambda, \sigma}$ for different real values, according to \eqref{eq:Gderivates}, for two different interval durations to be estimated.

\begin{table}[h]
	\centering
	\caption[Simulated hyperparameters estimation]{Simulated hyperparameters estimation}
	\label{tab:hypEst}
	\begin{tabular}{c|cc|cc|cc}
		\hline
		{Real values} & $\lambda$ = 0.65 & $\sigma$ = 0.45 & $\lambda$ = 0.65 & $\sigma$ = 0.2 & $\lambda$ = 0.3 & $\sigma$ = 0.45 \\ \hline
		{20 s} & 0.66 & 0.41 & 0.61 & 0.26 & 0.41 & 0.41 \\
		\hline
		{10 s} & 0.56 & 0.46 & 0.71 & 0.21 & 0.26 & 0.46 \\
	\end{tabular}
\end{table}

The results show a similarity between the statistical properties of the original processes and the estimated ones, from which we can conclude that from an unknown process we should be able to calculate the values of the hyperparameters that better represent its statistical model.

\subsection{Evaluation of the performance of the framework}
The choice of the interval duration for the experiment came from a 50\% of probability of being short, and 50\% of being long. Within each cathegory, there was a uniform probability of choosing any of the value. 

Table \ref{tab:RLtask} shows the setup of the RL time-dependent task. The value of the intersection between a row, $s_{t}$, and a column, $a_{t}$, indicates the state $s_{t+1}$ for which the agent goes in the next timestep. The value $N$ means that the agent gets a negative reward, and $Y$ a positive one, and in both cases the experiment ends. In all the others cases the reward is neutral, $r_{t}=0$.

The performance of the agent in each episode is evaluated according to how close it got to one of the steps in Table \ref{tab:points}, indicating how well the agent performed in that episode. 

\begin{table}[h]
	\centering
	\caption{Setup of the time-dependent task}
	\label{tab:RLtask}
	\begin{tabular}{|c|cccccc|}
		\cline{4-7}
		\multicolumn{3}{l}{\multirow{2}{*}{}} & \multicolumn{4}{|c|}{Action} \\ \cline{4-7}
		\multicolumn{3}{l|}{} & Start & Wait & Short & Long \\ \cline{1-7} 
		\multirow{3}{*}{\rotatebox[origin=c]{90}{State}} & 0) & \multicolumn{1}{l|}{Init} & 1) & 0) & N & N \\
		& 1) & \multicolumn{1}{l|}{Sound} & N & 1) or 2) & N or Y & N or Y \\
		& 2) & \multicolumn{1}{l|}{Interval} & N & 1) or 2) & N & N \\ \cline{1-7}
	\end{tabular}
\end{table}

\begin{table}[h]
	\centering
	\caption{The number of points indicates the performance of the agent in that episode}
	\begin{tabular}{|c|l|}
		\hline
		\textbf{Points} & \textbf{Description}  \\ \hline
0 & Did not press the \textit{Start} button.\\ \hline
1 & Pressed the \textit{Start} button.\\ \hline
2 & Did \textit{Wait} after hearing the first tone.\\ \hline
3 & Did \textit{Wait} until hearing the second tone.\\ \hline
4 & Presses the correct button according to the length of the interval.\\ \hline
	\end{tabular}
	\label{tab:points}
\end{table}

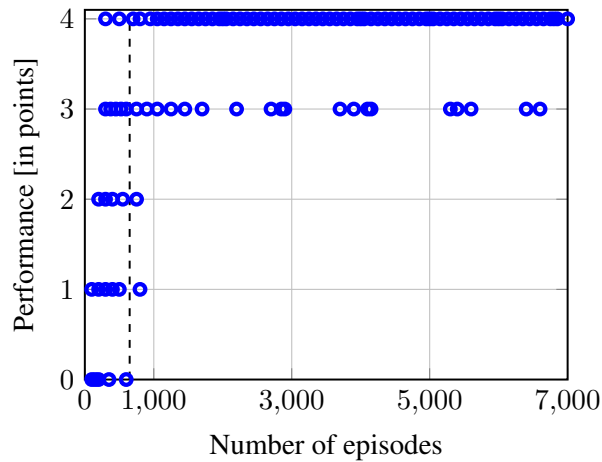
\begin{figure}[h]
  \centering
\begin{tikzpicture}
\begin{axis}[%
        xmin=0,
        xmax=7000,
        ymin=0,
        ymax=4.1,
        height=6.5cm,
        width=8cm,
        grid,
        xtick={0,1000,3000, 5000, 7000},
        ytick={0, 1,2,3,4},
        ylabel={Performance [in points]},
        xlabel={Number of episodes},
		grid,
		scatter/classes={%
   	 	a={mark=o,draw=blue, ultra thick}}]
\addplot[scatter,only marks,%
    scatter src=explicit symbolic]%
table[meta=label] {
x y label

300 4 a
500 4 a
700 4 a
800 4 a
950 4 a
2000 4 a
4000 4 a
5000 4 a
6000 4 a
7000 4 a
1050 4 a
1150 4 a
1250 4 a
1350 4 a
1450 4 a
1550 4 a
1650 4 a
1750 4 a
1850 4 a
1950 4 a
2050 4 a
2150 4 a
2250 4 a
2350 4 a
2450 4 a
2550 4 a
2650 4 a
2750 4 a
2850 4 a
2950 4 a
3050 4 a
3150 4 a
3250 4 a
3350 4 a
3450 4 a
3550 4 a
3650 4 a
3750 4 a
3850 4 a
3950 4 a
4050 4 a
4150 4 a
4250 4 a
4350 4 a
4450 4 a
4550 4 a
4650 4 a
4750 4 a
4850 4 a
4950 4 a
5050 4 a
5150 4 a
5250 4 a
5350 4 a
5450 4 a
5550 4 a
5650 4 a
5750 4 a
5850 4 a
5950 4 a
6050 4 a
6150 4 a
6250 4 a
6350 4 a
6450 4 a
6550 4 a
6650 4 a
6750 4 a
6850 4 a
6800 4 a

100 0 a
150 0 a
200 0 a
350 0 a
600 0 a

100 1 a
200 1 a
300 1 a
400 1 a
500 1 a
800 1 a

200 2 a
300 2 a
400 2 a
550 2 a
750 2 a

300 3 a
375 3 a
450 3 a
525 3 a
600 3 a
750 3 a
900 3 a
1050 3 a
1250 3 a
1450 3 a
1700 3 a
2200 3 a
2700 3 a
2850 3 a
2900 3 a
3700 3 a
3900 3 a
4100 3 a
4150 3 a
5300 3 a
5400 3 a
5600 3 a
6400 3 a
6600 3 a

    };
    \draw [dashed] (650,0) -- (650,4);
\end{axis}
\end{tikzpicture}
  \caption{TD learning with Microstimuli for a temporal discrimination task. The maximum interval was 8 timesteps.}
\label{fig:MS_8}
\end{figure}

Figure \ref{fig:MS_8} shows the evolution of the number of points achieved thoughout multiple episodes.
After around 1000 episodes, the agent already knows the correct sequence of actions to perform as to maximize the amount of reward. However, it can be seen that the algorithm does not converge, being that the episodes where it gets less than 4 points mainly correspond to those where the interval was in the boudary between short and long, as shown in Figure \ref{fig:MS_8a}.
The vertical dashed line shows the episode from which on the exploration decay parameter was $\epsilon < 0.01$.
%\begin{figure}[h]
%  \centering
%  \includegraphics[width=0.4\linewidth]{Figures/MS_8.png}
%  \caption{TD learning with Microstimuli for a temporal discrimination task. The maximum interval was 8 timesteps.}
%  \label{fig:MS_8}
%\end{figure}%

\subsection{Reward-prediction error}

\begin{figure}[h]
  \centering
  \includegraphics[width=0.8\linewidth]{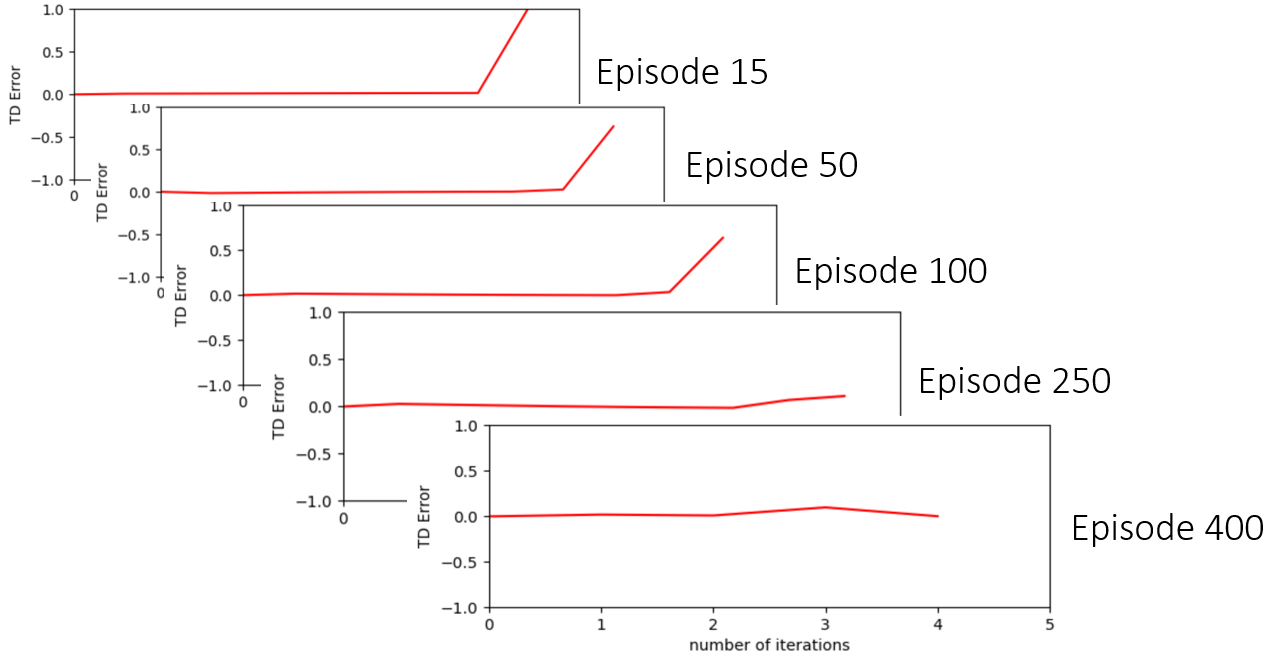}
  \caption{Evolution of the temporal-difference error, $\delta$, with the episodes.}
  \label{fig:TDerror}
\end{figure}

Figure \ref{fig:TDerror} shows the evolution of the temporal-difference error over 400 episodes. It represents that, as learning occurs, the difference between the expected and the received reward decreases more and more, which means that the agent starts to predict the reward. This presents an important confirmation of the desired performance of the algorithm since it replicates the behavior of dopamine neurons, which show a similar performance (e.g., see \citet{glimcher2011}).

%We consider the update rule:
%\begin{equation}
%e_{t}(s,a)=\begin{cases}
%1+ \gamma \Lambda e_{t-1}(s,a) , & \text{if $s=s_{t}, a= a_{t}, Q_{t-1}(s_{t},a_{t}) = \underset{a}{\mathrm{\max }} Q_{t-1}(s_{t}, a)$ }\\
%0, & \text{if $Q_{t-1}(s_{t},a_{t}) \neq \underset{a}{\mathrm{\max }} Q_{t-1}(s_{t}, a)$} \\
%\gamma \Lambda e_{t-1}(s,a), & \text{otherwise}
%\end{cases},
%\label{eq:etraces}
%\end{equation}
%where $\gamma$ is the discount rate and $\Lambda$ the decay parameter that determines the plasticity window of recent stimuli.

%
%\begin{table}[]
%	\centering
%	\caption[Average and standard deviation of the time estimates as time increases]{Average and standard deviation of the time estimates as time increases. The first row shows the real value of time, the second the average estimated and the third the standard deviation of the measures, all in {[}s{]}. The last row shows the error percentage, in {[}\%{]}.}
%	\label{tab:Average ans}
%	\begin{tabular}{l|llllllllllllll}
%		\hline
%		Real & 2 & 3 & 5 & 7 & 9 & 10 & 12 & 14 & 15 & 17 & 20 & 22 & 25 & 30 \\ \hline
%		Av. & 2.15 & 3.28 & 4.45 & 6.45 & 8.50 & 12.4 & 13.9 & 15.6 & 16.5 & 19.3 & 21.9 & 25.6 & 31.3 & 31.3 \\
%		S.d. & 1.13 & 1.69 & 2.17 & 2.41 & 3.68 & 4.1 & 5.37 & 5.9 & 6.6 & 8.4 & 9.3 & 10.6 & 10.2 & 12.1 \\
%		Error& 7.5 & 9.4 & 11.0 & 7.8 & 5.5 & 4.0 & 2.9 & 1.1 & 3.8 & 2.7 & 3.5 & 0.6 & 2.2 & 4.2
%	\end{tabular}
%\end{table}

% Acknowledgments---Will not appear in anonymized version
\acks{This work was partially supported by NewLEADS, the Swedish Research
Council, the Wallenberg AI, Autonomous Systems and Software
Program (WASP), and the FCT project [UID/EEA/50009/2019].}
%
%\begin{figure}[h]
%    \centering
%\includegraphics[width=0.4\textwidth]{Figures/EstMean2.png}
%\caption{\label{fig:time_estimate}Estimated $\tau$ for each real $\tau$.}
%\end{figure}
%
%\begin{figure}[h]
%    \centering
%    \subfigure[Intervals incorrectly classified based on their corresponding duration.]{
%        \includegraphics[width=0.4\linewidth]{Figures/MS_16a.png}
%        \label{fig:MS_8aa}} 
%    \subfigure[Probability of an interval being classified as short or long.]{
%        \includegraphics[width=0.4\linewidth]{Figures/MS_16b.png}
%        \label{fig:MS_8bb}}
%    \caption{The maximum interval duration in both was 16 timesteps.}
%\end{figure}

% External bibliography database file in the BibTeX format
%\bibliographystyle{unsrt}
\bibliography{L4DCRef.bib} % file "Thesis_bib_DB.bib"

% citep - in brackets
% citet - in text

\end{document}